\documentclass[prd,floatfix,showpacs,twocolumn]{revtex4}
\usepackage{epsfig}

\newcommand{\eq}[1]{Eq.~(\ref{#1})}

\newcommand{\be}{\begin{equation}}
\newcommand{\ee}{\end{equation}}
\newcommand{\bea}{\begin{eqnarray}}
\newcommand{\eea}{\end{eqnarray}}
\newcommand{\ben}{\begin{eqnarray*}}
\newcommand{\een}{\end{eqnarray*}}

\newcommand{\DS}{Dyson--Schwinger }

\newcommand{\al}{\alpha}

\newcommand{\G}{\Gamma}
\newcommand{\de}{\delta}

\newcommand{\si}{\sigma}

\newcommand{\ro}{\rho}
\newcommand{\la}{\lambda}

\newcommand{\ka}{\kappa}

\newcommand{\et}{\eta}
\newcommand{\ha}{\frac{1}{2}}
\newcommand{\pd}{\partial}

\renewcommand{\th}{\theta}
\newcommand{\cd}{{\cal D}}
\newcommand{\cs}{{\cal S}}
\renewcommand{\div}{\vec{\nabla}}

\newcommand{\s}[2]{{#1}\!\cdot\!{#2}}
\newcommand{\ov}[1]{\overline{#1}}
\newcommand{\dk}[1]{\,\,\,\raisebox{-0.4ex}{\large $\bar{}$}\!\!d\,{#1}\,}
\newcommand{\dx}[1]{d^4{#1}\,}

\newcommand{\ev}[1]{<\!\!{#1}\!\!>}
\begin{document}
\title{Completing Continuum Coulomb Gauge in the Functional Formalism}
\author{P.~Watson}
\author{H.~Reinhardt}
\affiliation{Institut f\"ur Theoretische Physik, Universit\"at T\"ubingen, 
Auf der Morgenstelle 14, D-72076 T\"ubingen, Deutschland}
\begin{abstract}
It is argued that within the continuum functional formalism, there is no 
need to supply a further (spatially independent) gauge constraint to 
complete the Coulomb gauge of Yang-Mills theory.  It is shown explicitly 
that a natural completion of the gauge-fixing leads to a contradiction with 
the perturbative renormalizability of the theory.
\end{abstract}
\maketitle
Consider Yang-Mills theory, invariant under the following (local) gauge 
transform characterized by the infinitesimal parameter $\th_x^a$ ($\si=A^0$):
\bea
\vec{A}_x^a&\rightarrow&\vec{A}_x^{\th a}=
\vec{A}_x^a+\frac{1}{g}\div_x\th_x^a-f^{acb}\vec{A}_x^c\th_x^b,\label{eq:gs}\\
\si_x^a&\rightarrow&\si_x^{\th a}=\si_x^a-\frac{1}{g}\pd_x^0\th_x^a
-f^{acb}\si_x^c\th_x^b.\label{eq:gt}
\eea
We are interested in Coulomb gauge, defined as the condition 
$\s{\div_x}{\vec{A}_x^a}=0$.  At the classical level it is clear that this 
condition restricts $\th_x^a$ to be independent of spatial argument 
$\vec{x}$ but which can be time-dependent or global.  Leaving aside the 
issue of global gauge fixing, the question addressed here is whether or not 
it is necessary to completely fix the local gauge; i.e., must we specify a 
time-dependent gauge condition in addition to the Coulomb gauge condition?  
It is shown that trying to complete the gauge-fixing leads to a 
contradiction and therefore that the system has in a sense `chosen' its own 
nontrivial completion of the gauge.

In the quantum, functional formulation of the theory, the central object of 
interest is the functional integral
\be
Z=\int\cd\Phi\exp{\left\{\imath\cs\right\}}
\ee
($\Phi$ denotes the collection of fields).  Because the action, $\cs$, (see 
\cite{Watson:2007vc} for our notation and conventions) is invariant under 
gauge transforms, we have to isolate the zero mode of the integral 
(generated by integration over the gauge group) and if we fix to Coulomb 
gauge using the Faddeev-Popov technique, then we have
\bea
Z&=&\int\cd\Phi\exp{\left\{\imath\cs+\imath\cs_{fp}\right\}},\nonumber\\
\cs_{fp}&=&\int d^4x\left[-\la^a\s{\vec{\nabla}}{\vec{A}^a}
-\ov{c}^a\s{\vec{\nabla}}{\left(\de^{ab}\div-gf^{acb}\vec{A}^c\right)}c^b
\right],\nonumber\\
\eea
where $\la^a$ is a Lagrange-multiplier field introduced to enforce the gauge 
condition, $\ov{c}^a$ and $c^b$ are Grassmann-valued ghost fields.  The 
functional integral above still contains zero modes corresponding to 
time-dependent gauge transforms (we do not consider the Gribov copies 
here).  If a further gauge-fixing condition is required to eliminate these 
zero modes, it must be spatially independent (so as not to interfere with 
Coulomb gauge itself) and we desire that it is local in the fields, such 
that functional techniques can be applied.  An obvious choice is
\be
F[\si]=\int d\vec{x}\si^a(x_0,\vec{x})=0.
\label{eq:gtfix}
\ee
The same condition exists when one considers the Coulomb gauge limit of 
interpolating (Landau-Coulomb) gauges in a finite volume and with periodic 
boundary conditions \cite{Baulieu:1998kx}.  Note that if we consider the 
(weaker) constraint
\be
\pd_x^0\int d\vec{x}\si^a(x_0,\vec{x})=0,
\ee
then since we have a single temporal dimension, this implies that
\be
\int d\vec{x}\si^a(x_0,\vec{x})=C
\ee
where $C$ is a constant.  However, under time-reversal, 
$\si^a(-x_0,\vec{x})=-\si^a(x_0,\vec{x})$, which forces $C=0$ and the 
condition, \eq{eq:gtfix}, above.  The form of the gauge-fixing condition, 
\eq{eq:gtfix}, has an immediate consequence in the functional formalism -- 
the Faddeev-Popov determinant generated is independent of the fields on the 
gauge-fixed hypersurface and is thus trivial.  To be specific, using the 
Faddeev-Popov technique we isolate the integration over the time-dependent 
gauge group by inserting the following identity into the functional integral:
\be
\openone=\int\cd\th\de(F[\si])\mbox{det}\left[M^{ba}(y_0,x_0)\right],
\ee
where
\bea
M^{ba}(y_0,x_0)&=&
\left.\frac{\de F[\si^{a\th}(x_0,\vec{x})]}{\de\th^b(y_0)}\right|_{F=0}
\nonumber\\
&=&\left[-\frac{1}{g}\de^{ba}\pd_x^0\de(y_0-x_0)\int d\vec{x}
\right.\nonumber\\&&\left.
-f^{acb}\de(y_0-x_0)\int d\vec{x}\si^c(x_0,\vec{x})\right]_{F=0}\nonumber\\
&=&-\frac{1}{g}\de^{ba}\pd_x^0\de(y_0-x_0)\int d\vec{x}.
\eea
In the above, the spatial integral of $\si$ is a number and vanishes due to 
the gauge condition $F=0$.  The remaining part of $\mbox{det}(M)$ is thus a 
pure number, independent of the fields and which can be incorporated into 
the normalization of the functional integral.  Clearly there will be no 
additional temporal Gribov ambiguity.  Our completely gauge-fixed functional 
integral now reads:
\be
Z=\int\cd\Phi\de\left(\int d\vec{x}\si_x^a\right)\exp{\left\{\imath\cs
+\imath\cs_{fp}\right\}}.
\ee
It is to be emphasized that the interaction content of the theory has not 
been modified by the extra gauge-fixing.  This means that the \DS equations 
will not change their form; what will change are the propagators and the 
effects will be seen at tree-level.  Thus, we may discard the interaction 
content of the theory from the discussion, save for one-loop integrals which 
will be considered later.

Let us then consider the generating functional of the theory by including 
source terms and restricting to at most quadratic terms in the action.  For 
definiteness, we express the $\de$-functional constraint as an integral over 
a new time-dependent Lagrange multiplier field, $\chi^a(x_0)$\footnote{In 
principle this is not the only method, but we do not expect the results of 
the study to be altered by using different technical constructs.}. We have
\bea
Z[J]&=&\int\cd\Phi\exp{\left\{\imath\cs+\imath\cs_s\right\}},\nonumber\\
\cs&=&\int\dx{x}\left\{-\ha A_i^a\left[\de_{ij}\pd_0^2-\de_{ij}\nabla^2
+\nabla_i\nabla_j\right]A_j^a
\right.\nonumber\\&&\left.
-\la^a\nabla_iA_i^a-\ov{c}^a\nabla^2c^a-\chi^a(x_0)\si^a
\right.\nonumber\\&&\left.
-A_i^a\pd_0\nabla_i\si^a-\ha\si^a\nabla^2\si^a\right\},\nonumber\\
\cs_s&=&\int\dx{x}\left\{\ro^a\si^a+J_i^aA_i^a+\ov{c}^a\et^a+\ov{\et}^ac^a
+\xi^a\la^a
\right.\nonumber\\&&\left.
+\ka_x^a\chi^a(x_0)\right\}.
\eea
The generating functional of connected Green's functions is $W[J]$, where 
$Z=e^W$ (in the context here, $J$ denotes a generic source).  Also defining 
the classical fields $\Phi_\al=\de W[J]/\de\imath J_\al$ we can construct 
the effective action, $\G$, as the Legendre transform of $W$:
\be
\G[\Phi]=W[J]-\imath J_\al\Phi_\al
\ee
(condensed index notation implies summation over all discrete indices and 
integration over all continuous arguments).  For notational convenience, we 
introduce a bracket notation to denote the functional derivatives of both 
$W$ and $\G$:
\be
\ev{\imath J_\al}=\frac{\de W}{\de\imath J_\al},\;\;\;\;\ev{\imath\Phi_\al}
=\frac{\de\G}{\de\imath\Phi_\al}.
\ee
We can now write down our tree-level equations for both proper and connected 
two-point functions using the techniques of 
\cite{Watson:2006yq,Watson:2007vc}.  In the case of the proper functions, 
this is more or less trivial -- for example, the Lagrange multiplier field 
$\la$ gives rise to the equation
\be
\ev{\imath\la_x^a}=-\nabla_{ix}A_{ix}^a
\ee
and all the further functional derivatives can be written down without 
ambiguity in either configuration or momentum space.  The case for the 
connected (propagator) two-point functions is far less clear.  The full set 
of equations reads:
\bea
\label{eq:ro0}\ro_x^a&=&
\pd_x^0\nabla_{ix}\ev{\imath J_{ix}^a}+\nabla_x^2\ev{\imath\ro_x^a}
+\ev{\imath\ka_x^a},\\
\label{eq:j0}J_{ix}^a&=&
\left[\de_{ij}\pd_{0x}^2-\de_{ij}\nabla_x^2+\nabla_{ix}\nabla_{jx}\right]
\ev{\imath J_{jx}^a}
\nonumber\\&&
+\pd_x^0\nabla_{ix}\ev{\imath\ro_x^a}-\nabla_{ix}\ev{\imath\xi_x^a},\\
\label{eq:xi0}\xi_x^a&=&\nabla_{ix}\ev{\imath J_{ix}^a},\\
\label{eq:ka0}\int d\vec{x}\ka_x^a&=&\int d\vec{x}\ev{\imath\ro_x^a},\\
\label{eq:et0}\et_x^a&=&\nabla_x^2\ev{\imath\ov{\et}_x^a}.
\eea

Let us begin with \eq{eq:xi0}.  The only non-zero functional derivative of 
the left-hand side is that with respect to $\imath\xi_y^b$, leading to
\be
-\imath\de^{ba}\de(y-x)=\nabla_{ix}\ev{\imath\xi_y^b\imath J_{ix}^a}.
\ee
The solution to this is written as
\be
\ev{\imath\xi_y^b\imath J_{ix}^a}=
\de^{ba}\int\dk{k}e^{-\imath k\cdot(y-x)}\frac{k_i}{\vec{k}^2}
\ee
where we recognize that when sources are set to zero, the function must be 
translationally invariant and be odd under the parity transform (it is a 
spatial vector).  This latter constraint necessarily precludes the 
possibility that there may be other (spatially independent) solutions to 
the homogeneous equation.  Indeed, the functional derivative of \eq{eq:xi0} 
with respect to $\imath\ro_y^b$ is just such a homogeneous equation:
\be
0=\nabla_{ix}\ev{\imath\ro_y^b\imath J_{ix}^a}.
\ee
Since $\ev{\imath\ro_y^b\imath J_{ix}^a}$ is a spatial vector, the function 
vanishes (as is clear if we Fourier transform into momentum space).  The 
same holds for $\ev{\imath\ka_y^b\imath J_{ix}^a}$: i.e.,
\be
\ev{\imath\ro_y^b\imath J_{ix}^a}=\ev{\imath\ka_y^b\imath J_{ix}^a}=0.
\ee
The spatial gluon propagator,
\be
\ev{\imath J_{jy}^b\imath J_{ix}^a}
=\int\dk{k}e^{-\imath k\cdot(y-x)}D_{AAji}^{ba}(k_0,\vec{k}),
\ee
is derived from the corresponding functional derivatives of equations 
(\ref{eq:j0}) and (\ref{eq:xi0}):
\bea
0&=&\int\dk{k}e^{-\imath k\cdot(y-x)}k_iD_{AAji}^{ba}(k_0,\vec{k}),\nonumber\\
0&=&\int\dk{k}e^{-\imath k\cdot(y-x)}\times
\nonumber\\&&
\left[(k_0^2-\vec{k}^2)D_{AAji}^{ba}(k_0,\vec{k})
-\imath\de^{ba}t_{ji}(\vec{k})\right]
\eea
($t_{ji}(\vec{k})=\de_{ji}-k_jk_i/\vec{k}^2$ is the transverse projector).  
The solution to this is
\be
D_{AAji}^{ba}(k_0,\vec{k})
=\de^{ba}t_{ji}(\vec{k})\frac{\imath}{(k_0^2-\vec{k}^2)}.
\ee

Now let us examine the ghost propagator stemming from \eq{eq:et0}.  Since 
the ghost field is Grassmann-valued whereas the propagators must be scalar, 
the ghost fields/sources must come in pairs.  Because the ghost fields 
anticommute, in the absence of sources the quantity 
$\ev{\imath\et_y^b\imath\et_x^a}$ must vanish -- we cannot construct any 
antisymmetric, color diagonal, scalar, translationally invariant function 
of invariants $(x_0-y_0)^2$ and $(\vec{x}-\vec{y})^2$.  Thus, the only 
functional derivative of \eq{eq:et0} that is of interest is
\be
\nabla_x^2\ev{\imath\ov{\et}_x^a\imath\et_y^b}=\imath\de^{ab}\de(x-y)
\ee
and the solution is
\be
D_c^{ab}(x_0-y_0,\vec{x}-\vec{y})
=-\de^{ab}\int\dk{k}e^{-\imath k\cdot(x-y)}\frac{\imath}{\vec{k}^2}.
\ee
In principle, we could add a homogeneous term $\sim\de(\vec{k})D(k_0^2)$ to 
the integrand above.  However, because the ghost propagator is connected to 
a ghost-gluon vertex with the factor $k_i$ in any loop integral 
\cite{Watson:2007vc}, the $\de$-function guarantees the situation whereby 
this term never appears in a calculation and we can disregard it.

Let us now turn to the remaining scalar propagators.  Since the $\la$ and 
$\chi$ fields are Lagrange-multiplier fields, propagators involving them 
will not contribute to any loop integral (they have no interaction term) and 
only the temporal gluon propagator, $\ev{\imath\ro_y^b\imath\ro_x^a}$, is of 
consequence.  We notice that \eq{eq:ka0} is integrated over $\vec{x}$ (a 
direct consequence of the fact that we must have a spatially independent 
second gauge condition) and will only determine the functions in momentum 
space at $\vec{k}=0$.  Indeed, we have that
\be
D_{\si\si}(k_0,\vec{k}=0)=0.
\label{eq:con}
\ee
This applies for all values of $k_0$, including the limit 
$k_0\rightarrow\infty$.  We are now led to a contradiction.  Since 
$D_{\si\si}$ is the only propagator that can cancel the well-known energy 
divergence of the ghost loop (see \cite{Watson:2007vc} for an explicit 
realization of this cancellation), it must have a finite part as 
$k_0\rightarrow\infty$, just as the ghost propagator, in order to effect the 
cancellation (we have explicitly shown that the mixed propagator, 
$D_{A\si}$, is zero and the spatial gluon propagator, $D_{AA}$, vanishes in 
this limit).  However, on dimensional grounds the above constraint, 
\eq{eq:con}, tells us that $D_{\si\si}$ must vanish -- in the absence of any 
external scale it can only behave as $k_0^{-2}(\vec{k}^2/k_0^2)^\mu$ for 
some positive power $\mu$ in this limit.  Moreover, if we try to determine 
$D_{\si\si}$ by solving the simultaneous set of equations generated from 
Eqs.~(\ref{eq:ro0}) and (\ref{eq:j0}) then we are led to the following 
(with the common color factors omitted):
\be
\left[\begin{array}{cccc}
k_0&-\imath&0&0\\
0&0&k_0&\imath\\
\vec{k}^2&0&-1&0\\
0&\vec{k}^2&0&1
\end{array}\right]
\left[\begin{array}{c}D_{\si\si}(k_0^2,\vec{k}^2)\\
k_0D_{\la\si}(k_0^2,\vec{k}^2)\\D_{\chi\si}(k_0^2,\vec{k}^2)\\
k_0D_{\chi\la}(k_0^2,\vec{k}^2)\end{array}\right]
=\left[\begin{array}{c}0\\0\\\imath\\k_0\end{array}\right].
\label{eq:mat}
\ee
The matrix in the above has a zero determinant (for all momenta) and the 
solution is determined only up to an unknown scalar function.  This function 
is only constrained by the requirement that it vanishes as $\vec{k}^2=0$ so 
as to agree with \eq{eq:con}.  The (physically meaningful) temporal gluon 
propagator is not determined, even at tree-level!  Thus, we have a situation 
whereby completely fixing the gauge has resulted in the energy divergences 
of the ghost loops not being canceled and the propagator content of the 
theory being ill-defined.

In summary, we have attempted to completely fix the Coulomb gauge by adding 
a further spatially independent gauge condition.  Whilst this extra 
condition seems justified (and even necessary) to deal with the zero modes 
in the functional formalism, its implementation has led to an explicit 
contradiction in defining the tree-level propagators of the theory.  There 
could be one of two reasons for this.  Since one is not familiar with such 
spatially independent constraints, the implementation here may be deficient 
in some way, although it is not clear how -- the condition \eq{eq:con} 
certainly appears a robust consequence of the gauge-fixing and the 
requirement of canceling the energy divergence of the ghost-loop is not 
ambiguous.  The more likely explanation is that the gauge-fixing condition 
contradicts the dynamics of the renormalizable quantum theory and in 
particular, Gau\ss' law.  In the functional approach, Gau\ss' law appears 
as the dynamical equation of motion for the $\si$-field and it is primarily 
this equation that leads to definition of the temporal gluon propagator, 
$D_{\si\si}$.  We are thus led to the conclusion that the construction of 
the functional formalism in Coulomb gauge implicitly `chooses' the remaining 
temporal gauge condition with the requirement of perturbative 
renormalizability so that a further constraint such as \eq{eq:gtfix} is not 
necessary.  Because in principle we should be able to choose any 
(reasonable) gauge-fixing condition, we can further say that if the 
condition given by \eq{eq:gtfix} is not allowed, then neither is any other 
condition, except that implicit condition `chosen' by the system itself.

Whilst we have used the second order formalism here, the same arguments will 
apply to Coulomb gauge in the first order formalism.  Formally, within the 
first order formalism, the system can be reduced to physical (spatially 
transverse gluon) degrees of freedom \cite{Zwanziger:1998ez,Watson:2006yq} 
and, on reflection, this would indeed seem to imply that there is no need 
for a further, spatially independent gauge constraint and in agreement with 
the conclusions here.  Further, the Gribov-Zwanziger confinement scenario 
\cite{Gribov:1977wm,Zwanziger:1998ez} in Coulomb gauge alludes to an 
infrared divergent temporal gluon propagator (from which a confining 
potential can be constructed) in contradistinction to condition \eq{eq:con}.

Given that the temporal gauge condition, \eq{eq:gtfix}, occurs in the 
interpolating gauge \cite{Baulieu:1998kx}, one might be tempted to infer 
from the results here that the Coulomb gauge end-point of interpolating 
gauge does not exist, or leads to different physical mechanisms when 
compared to Coulomb gauge.  This is not necessarily true and not our 
conclusion.  It is seen in the interpolating gauge lattice calculations of 
Ref.~\cite{Cucchieri:2007uj} that as one approaches the Coulomb gauge 
limit, an increasingly infrared (but still finite $|\vec{k}|$) enhanced, 
yet $\vec{k}=0$ vanishing temporal propagator emerges whilst the Coulomb 
gauge temporal propagator itself is infrared divergent.  Thus, it should be 
kept in mind that in discussing the perturbative propagators here, what 
matters physically are the integrals and combinations of the tree-level 
factors that form, for example, the nonperturbative propagators.  That the 
tree-level propagators, canceled energy divergences etc., have a different 
form in Coulomb gauge in distinction to interpolating gauges is not in 
itself either a drawback or a surprise -- quite tautologically, many 
different integrals have the same value.  Indeed, one may regard the 
differences between the internal constructions of Coulomb gauge and the 
interpolating gauge and how they still should result in the same observable 
physics as another fascinating piece of the puzzle to study.

\begin{acknowledgments}
The authors would like to thank R.~Alkofer for a critical reading of the 
manuscript.  They would also like to thank M.~Quandt and G.~Burgio for useful 
discussions.  This work has been supported by the Deutsche 
Forschungsgemeinschaft (DFG) under contracts no. DFG-Re856/6-1 and 
DFG-Re856/6-2.
\end{acknowledgments}

\end{document}